\documentclass[twocolumn,aps,showpacs,preprintnumbers,amsmath,amssymb,prl]{revtex4-2}

\usepackage{graphicx}
\usepackage{dcolumn}
\usepackage{bm}
\usepackage{color}
\usepackage{ulem}
\usepackage[colorlinks,bookmarks=true,citecolor=blue,linkcolor=blue,urlcolor=blue, breaklinks=true]{hyperref}

\begin{document}

\title{%
Spinon band flattening by its emergent gauge field {in quantum kagome ice}
}

\author{Masafumi Udagawa$^{1}$ and Roderich Moessner$^2$}%
\affiliation{%
$^1$Department of Physics, Gakushuin University, Mejiro, Toshima-ku, Tokyo 171-8588, Japan\\
$^2$Max-Planck-Institut f\"{u}r Physik komplexer Systeme, 01187 Dresden, Germany
}%

\date{\today}

\begin{abstract}
Fractional excitations provide a key to identifying  sought-after topological quantum spin liquid states in realistic materials. Their single-particle dynamics already presents  a challenging many-body problem on account of the coupling to their emergent gauge field.
Here, we study the spinon excitations of kagome ice, realized at the $2/3$ magnetization plateau of spin ice, by combining up-to-$63$-site exact diagonalization with an analytical state graph mapping. 
We find a macroscopically degenerate mode in the spinon spectrum. It  
originates from the destructive interference due to the interaction with surrounding gauge fields, a form of many-body caging.
We explicitly construct, and count, the concomitant many-body wave functions. 
Finally, we discuss the possible role of these flat modes in the  magnetization process of kagome antiferromagnets, in particular with regard to the asymmetric termination of the kagome ice magnetisation plateau.
\end{abstract}
                            
\maketitle
{\it Introduction}: 
The exotic properties of quantum spin liquid (QSL) have attracted considerable interests in condensed matter physics~\cite{ANDERSON1973153,balents2010balents,annurev:/content/journals/10.1146/annurev-conmatphys-031218-013401}.
Central to their rich phenomenology, in common with other topological phases of matter~\cite{moessner2021topological}, are the fractional excitations they host.

While there exist well-understood instances of quantum fractionalisation in dimension $d=1$~\cite{Lake:2010aa,Mourigal:2013aa,PhysRevLett.111.137205} and $d=2$~\cite{de-Picciotto:1997aa,PhysRevLett.79.2526,doi:10.1126/science.aaz5601,Nakamura:2020aa}, the experimental search in $d=3$ has been less conclusive. Nonetheless, there has been much recent progress, with various experiments on a family of compounds known collectively as quantum spin ices~\cite{Udagawa2021} providing strengthening evidence of (some form) of fractionalised quantum spin ice phase~\cite{Kimura:2013aa,Tang:2023aa,Uehara:2022aa,PhysRevB.94.165153,PhysRevLett.118.107206,PhysRevB.94.144415,Sibille:2018aa,Gao:2019aa,PhysRevLett.122.187201,Sibille:2020aa,Petit:2016aa,PhysRevLett.126.247201,tokiwa2016tokiwa,4qxy-l8pg,PhysRevB.111.155137,PhysRevB.106.094425,PhysRevX.12.021015,PhysRevB.108.054438,PhysRevX.15.021033,Gao:2025aa,Poree:2025aa}. In such a phase, the microscopic spins beget fractionalised deconfined spinons as emergent quasiparticles. These quasiparticles are charged under their concomitant emergent U(1) gauge field.  

As the experimental search for  quantum spin ice closes in, the following interconnected questions about the spinon dynamics naturally take centre-stage. Firstly, for any of the  
Hamiltonians for quantum spin ice models and materials, what quantitative predictions can be confronted with experiment? Secondly, what is the most appropriate framework for describing the fractionalised particles and the gauge field they are charged under? And thirdly, what are the qualitatively most striking features and phenomena to be found for this constellation of degrees of freedom?

Indeed, quantum spin ice differs in a number of important aspects from the familiar QED of high-energy physics: its fine-structure constant more than an order of magnitude larger \cite{PhysRevLett.127.117205} than the `usual' $\alpha=1/137$; it resides on the frustrated pyrochlore lattice, and the speed of the emergent charges is parametrically in excess of the speed of emergent light.  In addition, the gauge flux takes on a very simple form -- it is just the spin field itself -- so that its  dynamics can be easily described in terms of simple spin or loop flip operations.  

Such questions are of broad interest in the physics of strongly correlated electrons and well beyond. They have in particular played an important role in the description of high-temperature cuprate superconductors, where fractionalised quasiparticles coupled to an emergent gauge fields were one of the central conceptual innovations~\cite{BASKARAN1993853}, and whose description continues to provide a formidable challenge.

We pick up the above threads by considering perhaps the most 
direct signature of fractional excitations, namely the broad continuum observed in spectroscopic measurements. This generically arises because fractionalised particles never get created singly, and therefore,  scattering process involve many-particle continua. Crucially, such continua can still have characteristic features. 

For example, in quantum spin ice, a spectral-edge discontinuity is predicted as a consequence of highly-constrained dynamics of fractional magnetic monopoles~\cite{PhysRevLett.122.117201}.
Another edge discontinuity has been proposed based on the interaction between the magnetic monopoles and emergent photons~\cite{PhysRevLett.124.097204}.
Such theories  compared reasonably well with the recent inelastic neutron scattering experiments on Ce$_2$Sn$_2$O$_7$~\cite{Poree:2025aa}

We find that a particularly interesting setting is provided by  kagome ice, which is based on a macroscopically degenerate ground state of the $2/3$ magnetization plateau of spin ice in  [111] magnetic field~\cite{Matsuhira:2002aa,doi:10.1143/JPSJ.71.2365,PhysRevB.68.064411}. 
Its quantum version supports two species of fractional excitations, which we will refer to as spinon and triplon.
The {\it short-time/distance} behaviour of the spinons is not only very striking but also  elegantly analysed: they form a macroscopically degenerate flat band on account of the interplay of their motion with the emergent gauge field.
This flat band results from the destructive interference due to the coupling with surrounding gauge fields. This is in analogy to the one-particle flat bands frequently discussed in the kagome and other frustrated lattices~\cite{Sutherland1986,Bergman2008,Rhim2021,Mizoguchi2019,Maimaiti2021,Essafi2017aa,Bilitewski2018,Nakai2022,PhysRevB.110.L201109,ksdp-p4pm}. We note that ideas from the physics of single-particle flat bands can be carried over to this gauge-coupled case provided one views the latter as an effective hopping problem on the state graph (an object which has gained much prominence in the context of many-body localisation~\cite{BASKO20061126}).

Consequently, despite their many-body nature, the low-energy model of quantum kagome ice enables us to construct their wave function explicitly as a superposition of minimally 14 real-space-basis states.
The extremely high density of states of the flat band should be  detectable through spectroscopic experiments, and might cause asymmetric collapse of magnetization plateau, as discussed in the Heisenberg antiferromagnets~\cite{doi:10.1143/JPSJ.79.053707,PhysRevB.102.094419}.
It could also have  implications for the 1/9-magnetisation plateau observed in related models.

In the remainder of this paper, we first introduce the model and its background before explicating the above items in detail.

{\it Model}:
The quantum XXZ model on the kagome lattice in a magnetic field, 
\begin{align}
&\mathcal{H} = \mathcal{H}_{\rm CSI} + \mathcal{H}_{\rm ex} + \mathcal{H}_{\rm Z}\nonumber\\
&= J_z\sum_{\langle i,j\rangle}S_i^zS_j^z + \frac{J_{\pm}}{2}(S_i^+S_j^- + S_i^-S_j^+) - h\sum_jS_j^z.
\label{eq:Hamiltonian}
\end{align}
$S^{\alpha}_j$ is a spin-$1/2$ operator.
The first term expresses the dominant antiferromagnetic ($J_z >0$) Ising interaction between neighboring spins, while the  second term introduces quantum dynamics.
We set $J_{\pm} >0$. The last term is a Zeeman coupling to an external magnetic field.

\begin{figure}[ht]
\begin{center}
\includegraphics[width=0.49\textwidth]{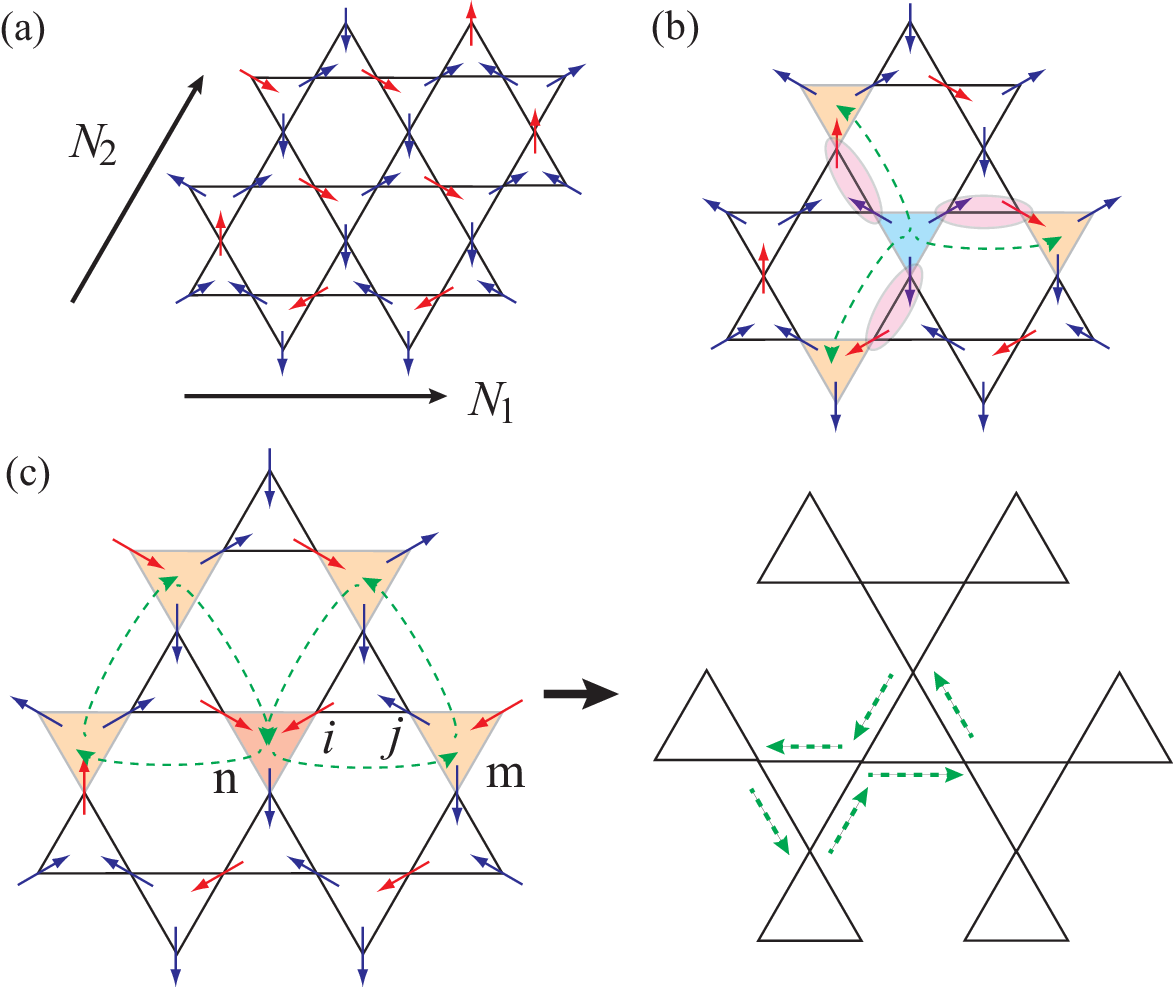}
\end{center}
\caption{\label{Fig1} 
(Color online) (a) {An example of kagome ice configuration.} (b) A triplon excitation on a downward triangle. A triplon can hop to neighboring three downward triangles with flipping pairs of spins as indicated by ovals. 
 (c) A spinon excitation on a downward triangle. A spinon can hop to neighboring four downward triangles.
The corresponding state graph locally takes the same structure as a triangular Husimi cactus. {Three successive hops of spinon around an upward triangle return all the spin configuration to the initial state, making a triangle in the state graph, as indicated by green dashed arrows.}}
\end{figure}

For the pure Ising model $J_{\pm}/J_z=0$, this realizes  classical kagome ice in a magnetic field range, $0<h<h_{\rm sat}\equiv2J_z$. 
The system shows a $1/3$ magnetization plateau, with two spins  $S_j^z=+\frac{1}{2}$ and one spin  $-\frac{1}{2}$ on each triangle. We will use a canted spin representation, Fig.~\ref{Fig1}, where
we orient the $+\frac{1}{2}$ spins from upward to downward triangles, and $-\frac{1}{2}$ spins vice versa. We can map these spin configurations to the kagome ice state realized at the $2/3$ magnetization plateau of spin ice [Fig.~\ref{Fig1} (a)].

Quantum dynamics to kagome ice is added via a small  $J_{\pm}$. Its effect is to lift the degeneracy of the classical kagome ice configurations, which  are connected by $\mathcal{H}_{\rm ex}$ only through third-order processes.
In contrast, the excitations are more strongly perturbed: they move at first-order in $\mathcal{H}_{\rm ex}$. 
We now analyse the quasiparticles dynamics in the perturbative region, $|J_{\pm}|\ll J_z$.

{\it Fractional excitations}:
Let us start with identifying the excitations of kagome ice.
One can violate the kagome ice rule in two ways.
Flipping an ``in" spin on a downward triangle  creates a 3-in upward triangle and a 3-out downward triangle, at a total energy cost of $2\Delta E_{\rm tri}=2J_z-h$, to  leading order of $\mathcal{O}(J_{\pm})$. Successive spin flips can separate these triangles, and isolate, say, the 3-out downward triangle [Fig.~\ref{Fig1} (b)].
We call  3-out downward (3-in upward) triangles  triplons; the excitation energy of a single triplon is  $\Delta E_{\rm tri}=J_z-\frac{h}{2}$.
At $h=h_{\rm sat}$, the triplons condense, marking the upper edge of the magnetization plateau.
$\mathcal{H}_{\rm ex}$ generates dynamics in first order of $J_{\pm}$ by flipping spins pairwise as shown in Fig.~\ref{Fig1} (b).
An isolated triplon has three pairs of flippable spins, and can hop to three nearest-neighbor downward triangles via $\mathcal{H}_{\rm ex}$.

The second violation of the kagome ice rule is what we call a spinon. A pair is generated by flipping an ``out" spin on a downward triangle, so that the ``wrong" kagome ice rule is imposed on the two triangles sharing this flipped spin: the upward triangle enters a 1-in 2-out and the downward triangle a 2-in 1-out configuration. Successive spin flips can again isolate a defect [Fig.~\ref{Fig1} (c)].
$\mathcal{H}_{\rm ex}$ endows the spinons with a  kinetic energy, $\varepsilon^{\rm sp}(J_{\pm})$, and the creation energy of single spinon may be written as $\Delta E_{\rm sp}=h+\varepsilon^{\rm sp}(J_{\pm})$. For $h\to0$, the spinon creation energy becomes negative at finite $h=h_{\rm c}$, corresponding to the band bottom of $\varepsilon^{\rm sp}(J_{\pm})$, which marks the lower edge of the magnetization plateau.

The spinon dynamics differs from that of the triplon: 
it can hop to four neighboring triangles.
Moreover, after three successive hops around a triangle, not only the spinon itself but the whole spin configuration returns to its initial configuration.

We find it useful to encode the spinon motion 
on a state graph, where each many-body state is represented by a vertex, and two vertices are connected with an edge if  $\mathcal{H}_{\rm ex}$ has a matrix element between the two corresponding  states. 
Due to the two features mentioned above, the state graph of the spinon consists of small triangular units: locally, it has  the  structure of a triangular Husimi cactus [Fig.~\ref{Fig1} (d)].

{\it Excitation spectrum}: By approximating the state graph as  a Husimi cactus,  we can obtain an analytical expression for the spectrum of the spinon excitations. This is approximate but accurate, as we will show below by comparison to exact diagonalisation.
This one-spinon density of states is obtained by solving the tight-binding model on the Husimi cactus via 
standard Green's function techniques:
\begin{eqnarray}
\rho^{(1)}(\varepsilon) = \frac{1}{3}\delta(\varepsilon+2) + \tilde{\rho}^{(1)}(\varepsilon)\ .
\label{eq:oneDOS}
\end{eqnarray}
While the second part yields a relatively featureless  continuum, $\tilde{\rho}^{(1)}(\varepsilon)=\frac{1}{\pi}\frac{\sqrt{((1+2\sqrt{2})-\varepsilon)(\varepsilon-(1-2\sqrt{2}))}}{(4-\varepsilon)(\varepsilon+2)}$, the first part is quite spectacular: it describes a flat mode at $\varepsilon=-2$ which hosts 1/3 of all the spectral weight. The corresponding states can be easily visualised as ``string modes" on the Husimi cactus:
the spinon wavefunction has equal weights of alternating signs along a string on the Husimi cactus. 
This mode thus results from the destructive interference due the local triangular structure of the network.

To make contact with the numerical results obtained as described below, and eventually experiment, it is useful to evaluate the  two-spinon DOS, as spinons cannot be created singly. It is given by
the convolution as $\rho^{(2)}(\varepsilon)=\int\ dx\rho^{(1)}(\varepsilon - x)\rho^{(1)}(x) $: 
\begin{align}
\rho^{(2)}(\varepsilon)=\frac{1}{9}\delta(\varepsilon+4) &+ \frac{2}{3}\tilde{\rho}^{(1)}(\varepsilon+2)\nonumber\\
 &+ \int\ dx\tilde{\rho}^{(1)}(\varepsilon - x)\tilde{\rho}^{(1)}(x).
\label{eq:twoDOS}
\end{align}
$\rho^{(2)}$ is also  highly asymmetric, with a steep singularity at the lower spectral edge.

\begin{figure}[ht]
\begin{center}
\includegraphics[width=0.5\textwidth]{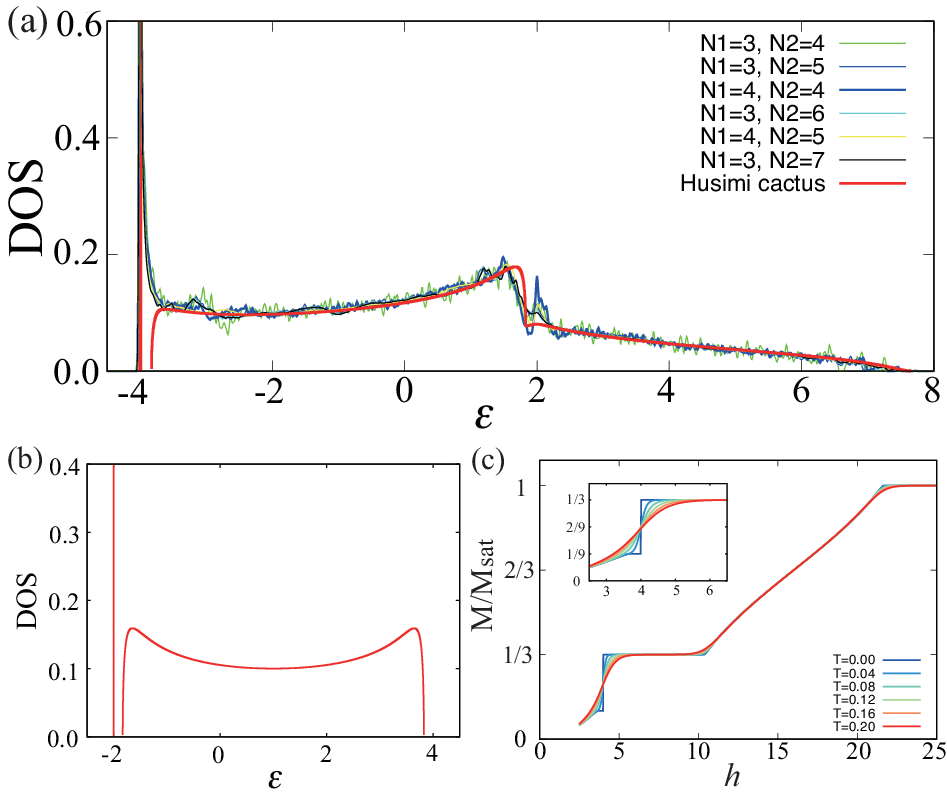}
\end{center}
\caption{\label{Fig2} 
(Color online) (a) Density of states in the two-spinon sector obtained by exact diagonalization of $N_{\rm spin}=3\times N_1\times N_2$-spin system, which are well fitted with $\rho^{(2)}(\varepsilon)$ obtained from Eq.~(\ref{eq:twoDOS}). (b) One-spinon density of states $\rho^{(1)}(\varepsilon)$ on the Husimi cactus, Eq.~(\ref{eq:oneDOS}). (c) Magnetization process based on Eq.~(\ref{eq:MagnetizationProcess}), taking $J_{z}=4J_{\pm}=8$, with several temperatures. The inset shows the magnetization curve around the $1/9$-plateau.}
\end{figure}

We find remarkable agreement between analytics and numerics, Fig.~\ref{Fig2} (a).  
The combination of two $\rho^{(1)}(\varepsilon)$ produces the delta-functional singularity of $\rho^{(2)}(\varepsilon)$ at $\varepsilon=-4$ with weight $1/3^2$.
Moreover, the convolution of singular and non-singular parts of $\rho^{(1)}(\varepsilon)$ lead to the kink-like structure of $\rho^{(2)}(\varepsilon)$ at $\varepsilon=2\sqrt{2}-1\simeq1.82$, which comes from the upper band edge of $\tilde{\rho}^{(1)}(\varepsilon+2)$.

We note that the numerical data is quite well finite-size converged. It was obtained from exact diagonalisation of $\mathcal{H}_{\rm eff}$, Eq.~(\ref{eq:kagheff}), on a kagome lattice of the $N_1\times N_2$ unit cells with periodic boundary conditions, on systems containing up to 63 spins. 
We apply the first-order degenerate perturbation theory to the Hamiltonian Eq.~(\ref{eq:Hamiltonian}) to derive the effective Hamiltonian,
\begin{eqnarray}
\mathcal{H}_{\rm eff} = \frac{J_{\pm}}{2}\sum_{\langle n,m\rangle}P(a^{\dag}_n\sigma^x_i\sigma^x_{j}a_{m} + {\rm H.c.})P.
\label{eq:kagheff}
\end{eqnarray}
Here, $P$ is the projection onto the Hilbert space defined by the classical kagome ice states in the presence of two spinons. $a^{\dag}_n$ is the creation operator of spinon, which hop from  triangle $m$ to $n$ by flipping the intervening spins at $i$ and $j$,
as shown in Fig.~\ref{Fig1} (c). We have used $\frac{J_{\pm}}{2}=1$ as a unit of energy.

Besides this remarkable consistency, there is one qualitative difference: the analytical result exhibits a gap between the delta peak and continuum. This gap is an artefact arising from the `infinite-dimensional' expander nature  the Husimi cactus shares with Bethe lattices in general~\cite{PhysRevB.80.144415}. It is now natural to ask to what extent the string mode on the Husimi cactus correctly captures the actual flat mode wavefunction.

This wavefunction turns out to be beautifully simple: the flat mode forms a closed loop of equal weights alternating in sign of the type familiar  
from tight-binding models on the kagome and pyrochlore lattices.
However, the crucial difference is that the flat mode is of many-body origin: the closed loop resides on the state graph, not in a real space! It is thus an instance of a many-body caged wavefunction{~\cite{benami2025manybodycagesdisorderfreeglassiness,tan2025interferencecagedquantummanybodyscars,nicolau2025fragmentationzeromodescollective,jonay2025localizedfockspacecages}}.

\begin{figure}[ht]
\begin{center}
\includegraphics[width=0.5\textwidth]{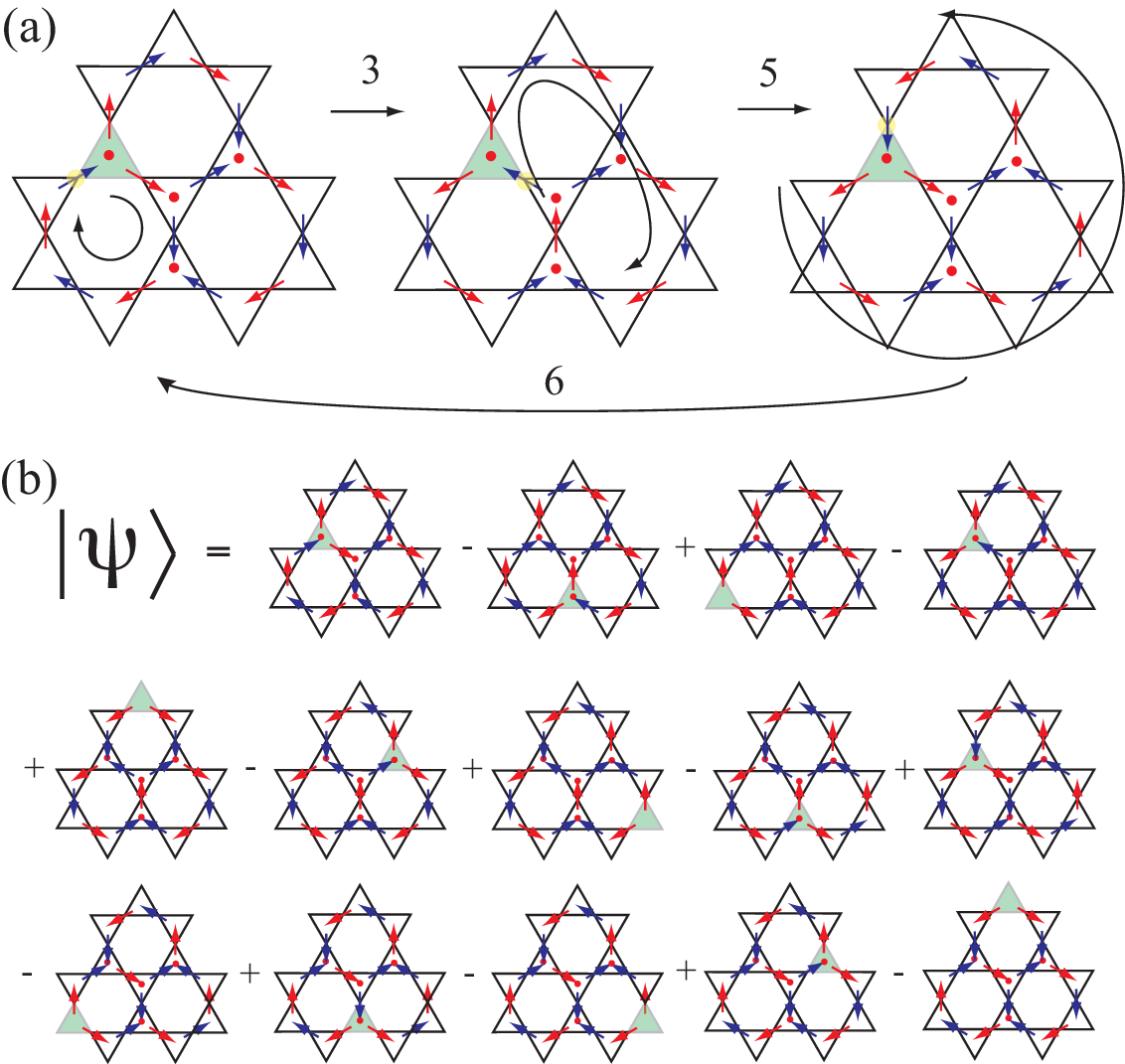}
\end{center}
\caption{\label{Fig3} 
(Color online) (a) A schematic figure of the procedure to construct a localized spinon flat mode. (b) The minimal localized flat mode as a superposition of 14 real space basis states.}
\end{figure}

Fig.~\ref{Fig3} (a) identifies the shortest such  closed loop on the state graph, where the spinon moves along a closed path, at the end of which all spins return to their original configuration. 
As a spinon hops to a neighboring triangle, it flips a pair of intervening spins; if the same bond is traversed twice by the spinon, 
the orientation of flipped spins is restored.
This ``two-stroke drawing" is concretely realised, Fig.~\ref{Fig3} (a), by first moving a spinon around a hexagon clockwise in three steps, then around two hexagons clockwise in five steps, then finally, around the three hexagons counterclockwise in six steps. 
The superposition with alternate signs of the 14 configurations involved  gives the expression of spatially localized flat mode [Fig.~\ref{Fig3} (b)].

The state graph description also gives an elegant explanation of the weight of the flat mode by appealing to a beautiful result obtained by considering  the state graph $\mathcal{G}$ of a single spinon as the so-called line graph of $G$, defined as follows. $\mathcal{G}$ is composed of triangles as the smallest unit; their centre-points define the sites of $G$, whose bonds/edges in turn are defined by the sites of $\mathcal{G}$.

Now, the {lattice Laplacian of} $\mathcal{G}$ involves $|\mathcal{G}| - |G|$ zero modes, while the other non-zero eigenvalues are exactly the same.
Since $|G| = \frac{2}{3}|\mathcal{G}|$, it means at least $1/3$ of the one-spinon eigenstates make up the flat mode. 
This accounts for the delta-functional peak has $1/3$ of the total spectral weight, as shown in Eq.~(\ref{eq:oneDOS}), and 
the degeneracy of  $(1/9)^\mathrm{th}$ of the two-spinon states, Eq.~(\ref{eq:twoDOS}).
The line-graph argument shows that the weight of $1/9$ of Eq.~(\ref{eq:twoDOS}) is not an artifact of Husimi cactus approximation.

{\it Magnetization process}: The flat spinon mode has direct implications for the magnetisation curve, in particular at the lower edge of the $1/3$ magnetization plateau of the Hamiltonian Eq.~(\ref{eq:Hamiltonian}).
If magnetic field is lowered to $h_{\rm c}=2J_{\pm}$, a macroscopic number of spinons appear, so that the magnetization plateau terminates with a discontinuous change of magnetization.
This magnetization jump is susceptible to thermal broadening, as the occupation rate of flat mode sensitively changes at finite temperatures.
This is in sharp contrast to the upper edge of plateau, where triplons with a conventional  dispersion effect a continuous change of magnetization. 

We demonstrate these robust consequences of the spinon flat modes on magnetization process in Fig.~\ref{Fig2}(c).
To this aim, as it is difficult to accurately consider the repulsion between spinons in the flat mode, we for convenience assume a Fermi distribution function $f(\varepsilon)$ for spinons and triplons to 
mimic their hardcore constraint, and describe the magnetization curve:
\begin{align}
\frac{M}{M_{\rm sat}} = \frac{1}{3} + \frac{2}{3}\int\ d\varepsilon\ \rho_{\rm tpl}^{(1)}(\varepsilon)&f(\varepsilon+J_z-\frac{h}{2})\nonumber\\
 &- \rho^{(1)}(\varepsilon)f(\varepsilon+\frac{h}{2}),
\label{eq:MagnetizationProcess}
\end{align}
where $\rho^{(1)}(\varepsilon)$ is derived from Eq.~(\ref{eq:oneDOS}), and $\rho_{\rm tpl}^{(1)}(\varepsilon)=\frac{3}{2\pi}\frac{\sqrt{8-\varepsilon^2}}{9-\varepsilon^2}$ is the triplon density of states.
With decreasing magnetic field, the magnetization jumps at $h=h_{\rm c}=2J_{\pm}$, where the flat mode  fills up.

This big magnetization jump thus serves as  evidence of the many-body flat mode specific to  quantum kagome ice.
It is remarkable that the  flat spinon mode of quantum kagome ice provides such a simple experimentally observable signature.

Following the jump, the magnetization shows a plateau at $\frac{M}{M_{\rm sat}}=\frac{1}{9}$ of  width $\delta h = \frac{3-2\sqrt{2}}{2}J_{\pm}$, corresponding to the spectral gap [Fig.~\ref{Fig2} (c) inset].
While the presence of gap itself follows from the Husimi cactus approximation of the state graph, it is tantalising to ask if it does give rise to a vestigial flat region following the magnetization jump, when the correct loop structure of state graph is taken into account. In this context,
it is interesting to point out that a $\frac{1}{9}$ plateau has been reported for the antiferromagetic Heisenberg model on the kagome lattice~\cite{Nishimoto:2013aa,Okuma:2019aa,PhysRevB.93.060407,doi:10.7566/JPSJ.93.123706}.
Moreover, {in addition to the $1/3$ plateau observed in kagome compounds~\cite{Kato:2024aa,PhysRevLett.114.227202,Okuma:2019aa,PhysRevB.102.104429,PhysRevB.100.174401,PhysRevLett.132.226701},} a glimpse of a $1/9$ plateau has recently been caught in {Y Kapellasites, YCu$_3$(OH(D))$_{6+x}$Br$_{2+x}$ ($x\sim0.5$)~\cite{Jeon:2024aa,PhysRevLett.132.226701}}.
Although those $1/9$ plateaus were reported for  systems with rather isotropic magnetic interactions, we note that the applied magnetic field at any rate removes the Heisenberg symmetry by defining a special direction, as does our choice of Ising axis.

Indeed, several other results of isotropic systems qualitatively resemble our picture based on spinons and triplons. 
Sakai and Nakano~\cite{doi:10.1143/JPSJ.79.053707} reported that the density of states of low-energy excitations seems to diverge at the lower edge of the plateau, in contrast to the conventional behavior at the upper edge.
Misawa and Yamaji~\cite{PhysRevB.102.094419} used finite-temperature Lanzcos method to show that the lower edge of the plateau is much more fragile at finite temperatures.
In particular, the fragility of the low-field edge of the plateau chimes with the extremely high density of states of the spinon flat mode.

In summary, our combined analytical and numerical analysis of  the spectrum of spinons in quantum kagome ice has yielded a quantitatively detailed and conceptually complete picture. This has provided a concrete instance of flat band physics on the state graph~\cite{benami2025manybodycagesdisorderfreeglassiness,tan2025interferencecagedquantummanybodyscars,nicolau2025fragmentationzeromodescollective,jonay2025localizedfockspacecages}, including exact wavefunctions as well as an evaluation of the weight of the band. These in turn  may account for the asymmetric termination of the magnetization plateau observed in  previous theoretical studies and may contribute to an understanding of new magnetization plateaux of kagome antiferromagnets more generally. A complete treatment of the magnetisation plateau however requires taking into account the interactions between spinons. This we would like to leave to a future study.

Finally, the state graph perspective has allowed us unusual analytic control over the problem of the joint dynamics of emergent fractionalised quasiparticles and their emergent gauge fields.  This provides a rare instance of an accurate implementation of the hard gauge constraint involving fractionalised quasiparticles, and represents a solution of the {\it many-body problem} defined by the motion of a {\it single} spinon.

\acknowledgments
 This work was supported by the JSPS KAKENHI (Nos. JP20H05655, JP22H01147, and JP23K22418), MEXT, Japan, and by the Deutsche Forschungsgemeinschaft 
 via the cluster of excellence ct.qmat (EXC 2147, project-id 390858490) and  SFB 1143 (Project-ID No. 247310070).

\bibliographystyle{apsrev4-2}
\bibliography{arXiv_manuscript.bbl}

\end{document}